# Spin-lattice Coupling and Magneto-dielectricity in $Ca_3Mn_2O_7$ Layered Perovskite

*Pooja Sahlot, Gaurav Sharma, Vasant Sathe, and Anand Mohan Awasthi**

UGC-DAE Consortium for Scientific Research, University Campus, Khandwa Road, Indore-452 001, India.
*E-mail: amawasthi@csr.res.in

**Abstract:** Dielectric study on $Ca_3Mn_2O_7$ features relaxor-like segmented dynamics below the antiferromagnetic ordering. Dipolar relaxations of different origin are spectrally resolved exhibiting distinct *H*-field alterations. This identifies their allegiance to different magnetic sub-phases and establishes dual coupling of electrical, magnetic, and structural degrees of freedom. Further, strong spin-lattice coupling has been affirmed with Raman spectroscopy across the magnetic ordering. Short-range electrical correlations collaterally cause measurable harmonic dielectric response in the system. The $\chi_3^e$-susceptibility signal yields genuine harmonic magneto-dielectricity, consistent with but exhibiting two orders of magnitude larger *H*-field effect, vis-à-vis that obtained in the fundamental dielectric constant $\varepsilon'$.

## Introduction

Interesting evolution of magnetic and electrical configurations has been realized in layered Ruddlesden-Popper (R-P) compounds, with inter-coupled structural, magnetic, and electric properties. R-P series compounds are represented as $A_{n+1}B_nO_{3n+1}$ for 2+ valence state 'A' cations and 4+ valence 'B' ions, with C usually being oxygen (2- valence).[1] Structurally, they reveal themselves as integral-*n* $ABO_3$ perovskite blocks, with inserted AO sheets in-between, along the *c*-axis. Here layered structure has open doors to many interesting properties. $La_{1.4}Sr_{1.6}Mn_2O_7$ has been shown to possess inter-planar tunneling induced low-field magneto-resistance.[2] Development of ferromagnetic clustering in the long range antiferromagnetic (AFM) ordering has been reported in *n* =3 R-P compound $La_{3-x}Sr_{1+3x}Mn_3O_{10}$.[3] Further, in R-P compounds the ferroelectric mechanism has been described via combination of two octahedral distortion patterns namely “hybrid improper ferroelectricity”.[4] Hybrid improper ferroelectricity (HIF) supported by first-principles calculation has been evident in *n* =2 R-P compound; $Ca_3(Ti,Mn)_2O_7$.[5,6] Thus compared to simple-perovskites these distortions offer scope for magneto-electric (ME) coupling and multiferroicity in these layered structures; since unfulfilled *d*-orbital accommodated in the oxygen octahedra favors (anti)ferromagnetism, which relates profoundly with the metal-oxygen-metal bond angles.[7]

$Ca_3Mn_2O_7$ R-P system exhibits structural transition from I4/mmm tetragonal phase to inversion-symmetry-breaking Cmc21 orthorhombic phase, with introduction of octahedral rotation in (001) plane, along with the octahedral tilt about [100] axis.[8] The structural transition covers a wide (300-600K) temperature range**,** with coexistent tetragonal and orthorhombic phases.[9] In pure orthorhombic phase, relaxor behaviour has been suggested below the room temperature.[9] Because of the high leakage-current at high temperatures in the orthorhombic phase**,** there have been difficulties in experimentally ascertaining ferroelectricity in the system.[7] Upon cooling, dielectric losses get reduced and ferroelectricity with measurable polarization is detected below 60K, which can be further explored at lower temperatures.[9] Here, emphasis on *n* =2 R-P manganite $Ca_3Mn_2O_7$ is due its expected candidacy for ME coupling.[4] Complementarily to ferroelectricity, octahedral tilting supports weak ferromagnetism and octahedral rotation induces magneto-electricity in this system.[10] Electric field tunability of these octahedral rotations is expected to enhance the ME effect. From first-principles calculations, Benedek and Fennie have reported G-type antiferromagnetic (AFM) ground state for the system, with a net perpendicular spin-canted moment via spin-orbit (S-O) interactions.[4] Dzyaloshinskii’s criteria explain the canted moment via the oxygen-octahedral tilt distortion. In $Ca_3Mn_2O_7$ system, antiferromagnetic transition with Nèel temperature 123K, along with the emergence of weak ferromagnetism (WFM) below 110K, have been reported previously.[11] Dzyaloshinskii-Moriya (D-M) interaction favored WFM clusters' formation in the AFM-matrix, inducing an exchange-bias (EB), was characterized in detail.[11,12] Here, we present the linear/non-linear dielectric properties of $Ca_3Mn_2O_7$, and their correlation with magnetic-evolution/phase-mixing is compared & distinguished. Here non-linear dielectric measurements compliment the understanding of electrical structure from linear measurements, giving more profound effect of change in magnetic-configuration on dipolar-correlations compared to latter. Magneto-dielectricity (linear and non-linear) is explored under applied magnetic field, across a sufficiently wide temperature range covering the AFM ordering at $T_N$ =123K.

**Experimental Details**

Single phase polycrystalline $Ca_3Mn_2O_7$ was synthesized at 1300°C by conventional solid state synthesis technique. Structural characterization by powder XRD using Bruker D8 advance diffractometer with Cu-$K_\alpha$ X-ray radiation ($\lambda$ =1.5405 Å) confirms orthorhombic structure with lattice parameters obtained as, $a$ =19.40(8) Å, $b$ =5.24(3) Å, and $c$ =5.25(2) Å, as reported previously.[11] Raman scattering measurements on $Ca_3Mn_2O_7$ system have been performed with HR800 Jobin-Yvon spectrometer, in the temperature range of 5K to 300K, using He-Ne laser of wavelength 632.8 nm. Low temperature fundamental dielectric response and magneto-dielectricity along with second harmonic measurements have been performed in parallel-plate capacitor configuration, from 200K down to 7.5K in the frequency range of 50 Hz to 300 kHz, using NovoControl Alpha-A Broadband Impedance Analyzer and Oxford NanoSystems Integra 9T magnet-cryostat.

**Results**

***Raman Spectroscopy*:** To explore the spin-lattice coupling in the system via study of vibrational modes, Raman scattering measurements have been performed. Predicted Raman active optical modes for the system with Cmc21 space group are $18A_1+17A_2+16B_1+18B_2$. The Raman spectrum depicting phonon modes for the sample are shown in **figure 1(a)**, in the temperature range of 5K to 300K. The two most prominent phonon modes have been recognized as $A_1^{(15)}$ and $A_1^{(17)}$ for the Cmc21 phase (=A21ma structure with '*c*'-long axis).[13] Lorentzian peak-fit analysis (**figure 1(b) and 1(c)**) has been carried out for the respective modes, to obtain thermal evolution of the phonon modes. Peak-characteristics follow the classical behaviour; frequency-shifts decrease and line-widths broaden with increase in temperature. This is accompanied by sharp changes in the frequency-shifts at ~105K; shown in **figure 1(d)** (left-pane) for $A_1^{(15)}$ mode and in **figure 1(e)** (left-pane) for $A_1^{(17)}$ mode, which independently confirm the spin-lattice coupling in the system. Line-widths of the respective modes have also been studied, which relates to the decay/lifetime of the phonons involved. Concurrent kinks in the line-widths again at ~105K, shown in figure 1(d) (right-pane) and figure 1(e) (rightpane) for $A_1^{(15)}$ and $A_1^{(17)}$ modes respectively, establish the spin-lattice coupling.

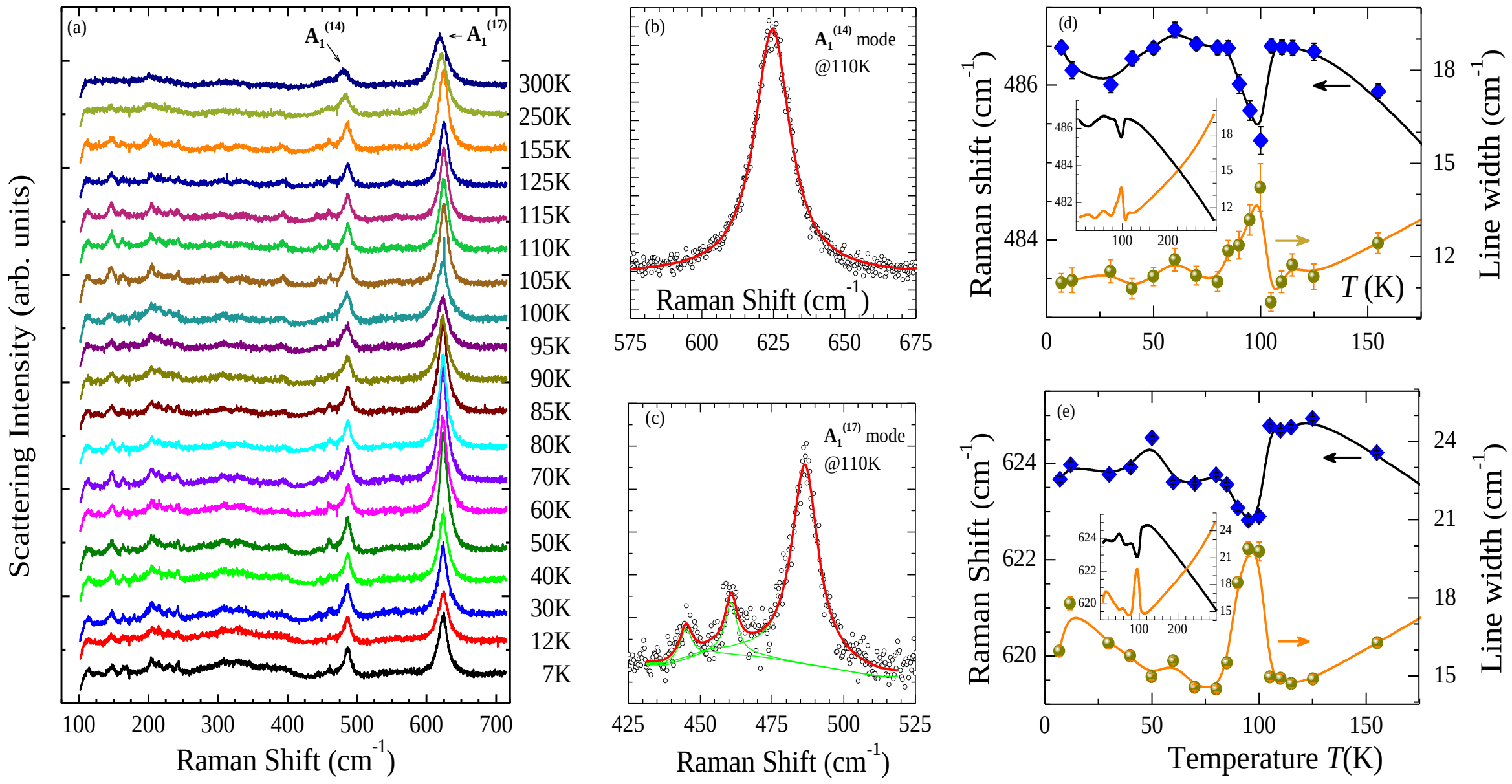

**Figure 1.** (a) Obtained Raman spectra from room temprature down to 7K. (b), (c) Lorentzian peak-fit for $A_1^{(15)}$ mode and $A_1^{(17)}$ mode repectively. (d) Temperature dependence of Raman shift (left-pane) and line-width (rigth-pane) obtained for $A_1^{(15)}$ mode. (e): Temperature dependence of Raman shift (left-pane) and line-width (rigth-pane) obtained for $A_1^{(17)}$ mode.

Signals corresponding to the co-existing WFM-phase (below 110K) in the AFM-matrix phase ($T_N$ =123K) are thus clearly revealed in the Raman study. Here, $A_1^{(15)}$ phonon mode mainly involves atomic motion of O1(-*z*), O3(*z*), and O4(*z*), along the *z*-axis (being the *a*-axis here) with in phase motion of O3 and O4. $A_1^{(17)}$ phonon mode characterizes in phase motion of O3(*z*), O4(*z*) atoms and stretching type motion of O1(*z*) atom. The change in the phonon modes across WFM further establishes the effective contribution from O1-atomic motion along the *a*-axis, in concurrence with the decrease in Mn-O1-Mn bond angle, obtained in the SXRD data analysis.[10] This explains that the Mn-$O_6$ octahedra-distortion related changes in the Mn-O-Mn bond angles induce changes in the magnetic configuration. Here, Raman study establishes that the WFM is associated with the inter-octahedral changes connected via O1-atom along the a-axis. This suggests low dimensionality of the WFM phase in this system.

***Dielectric Spectroscopy*: Figure 2(a)** presents the dielectric constant $\varepsilon'_f(T)$ of $Ca_3Mn_2O_7$ at various frequencies. Here, sharp frequency-dependent step-change in $\varepsilon'(T)$ is observed. At high-*T*'s, relaxation times $\tau(T)$ are small enough that even at higher radio frequencies (RF) condition $2\pi f \tau(T) < 1$ applies; resulting in a rather small frequency-dependence of $\varepsilon'_f(T)$, as observed. Cooling in the orthorhombic phase develops polar correlations; manifest here as step-like (thermally-activated) decrease of the dielectric constant. Thus-increased $\tau(T)$ leads to resonance of the measurement frequency and the response timescale of the correlated dipoles ($2\pi f \tau(T) =1$); which marks maximum slopes (inflexion-points) of $\varepsilon'_f(T)$, at successively lower frequencies upon cooling. Here, the correlations produce size-dispersed dipole-clusters, each responding optimally at different frequency ($\omega_p(T)$ ~1/cluster-size), which shows up in thermally activated dielectric spectra. For a given measurement frequency, super-resonance ($2\pi f \tau(T) >1$) condition upon further cooling causes the roll-off of $\varepsilon'(T)$ to lower values, for the same inertial reason as occurs for the $\varepsilon'(f)$-isotherms at high frequencies.

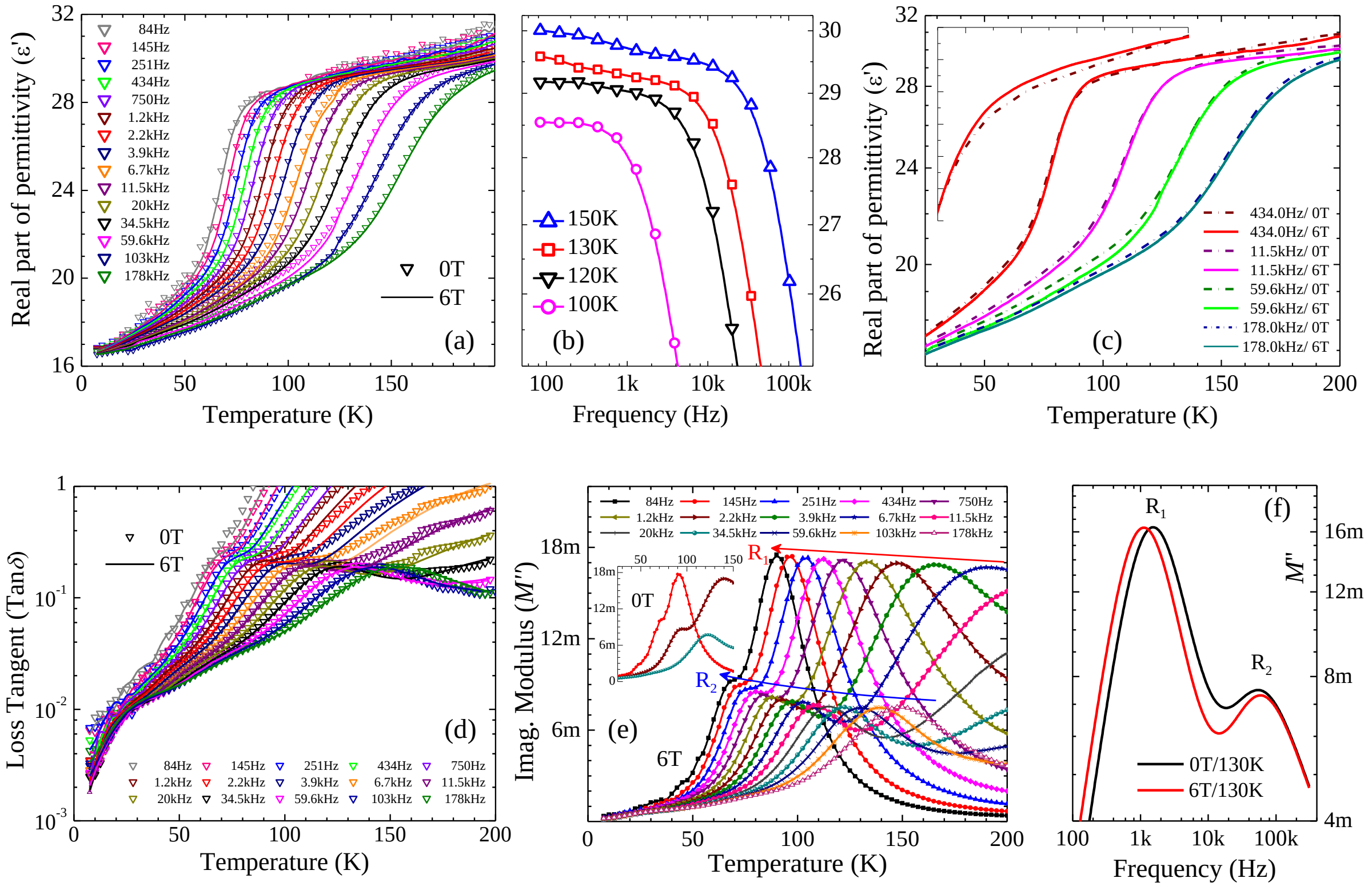

**Figure 2.** (a) Temperature dependent dielectric constant at different frequencies. (b) frequency dependence of dielectric constant at temperatures:100K, 120K, 130K & 150K. (c): $\varepsilon'(T)$ at selected frequencies under zero- and 6T magnetic field. (d): Temperature dependent loss-tangent at different frequencies, under zero- and 6T magnetic field. (e): Modulus $M''(T)$ isochrones under 6T applied $H$-field at various frequencies. Inset shows representative isochrones without the $H$-field. (f): frequency dependence of Modulus $M''(f)$ at 130K under 0T- and 6T-field.

In layered perovskites, FE-instability consists of O6-octahedron rotations, with negligible off-centering of the B-site cation. The associated energy-gain (due to lattice-strain) and energy-loss (due to dipole-dipole interaction) here rather settle in introducing antiferro-distortive (AFD) -like features over a temperature window, thereby suppressing the long-range FE-ordering in the system at a sharp transition temperature.[14,15] In the quantum paraelectric-like AFD systems, chemical substitution, application of electric field, or dimensional-strain can induce the FE.[16] In the layered $Ca_3Mn_2O_7$, no specific heat anomaly pertaining to a long-range FE transition is observed below the room temperature.[17]

Change in $\varepsilon'(T)$ produced under $H$-field is zoomed-on in **figure 2(c)**, at selected frequencies. At temperatures above the $\varepsilon'_f(T)$-step, $H$-field enhances correlations between the dipoles, evident from $\varepsilon'$(6T) < $\varepsilon'$(0T) (figure 2(c)). Also, for frequencies $f \geq 11.5$ kHz, at all temperatures $\varepsilon'_f(T)$ simply reduces under the field. At lower frequencies however, $H$-field alters the dielectric constant in both +ve and -ve sense over different temperature ranges, depending upon the extent of the dipolar interactions. At probing frequencies $f < 11.5$kHz, $H$-field mobilizes the larger/frozen dipolar clusters ($2\pi f \tau_{cl} > 1$; $\varepsilon'$(6T) > $\varepsilon'$(0T) over 80-130K) and consolidates/freezes the smaller ones ($2\pi f \tau_{cl} < 1$; $\varepsilon'$(6T) < $\varepsilon'$(0T) below 80K).

Associated with the step-like change in $\varepsilon'(T)$, relaxation peaks in loss tangent ($\tan\delta = \varepsilon''/\varepsilon'$) are observed, as shown in **figure 2(d)**. Increase in the peak temperature of $\tan\delta_T(T)$ isochrones with rise in frequency depicts thermally activated character. Here we emphasize low values of permittivity with seemingly frequency independent at low frequencies on cooling below 130K (**figure 2(b)**) makes the dielectricity observed for the system to be inter-grain response. Further $\tan\delta < 1$ for the relaxations excludes contributions from charge carriers for frequencies ≥ 100Hz.

For further insight, dielectric modulus ($M^*$) is observed, which reveals two sets of localized dipole-relaxations in the system.[18]

$$M^* = 1 \div \varepsilon^* = M' + iM'' \qquad (1)$$

$M''(T)$ isochrones shown in **figure 2(e)** reveal two relaxation processes in the system. **Figure 2(f)** depicts $M''(f)$ isotherm at 130K, here relaxation-$R_1$ well-developed at lower frequencies with higher modulus— persisting to higher frequencies with increase in temperature— and another relaxation-$R_2$ with comparatively low modulus well-formed at relatively higher frequencies. $R_1$-relaxation peaks shifts to lower frequency side with application of $H$-field while $R_2$-relaxation peak shifts to higher frequency side with *H*-field application. Our Havriliak-Negami analysis reveals, the corresponding peaks are not Lorentzian (broadening parameters $\alpha$, $\beta < 1$).[10]

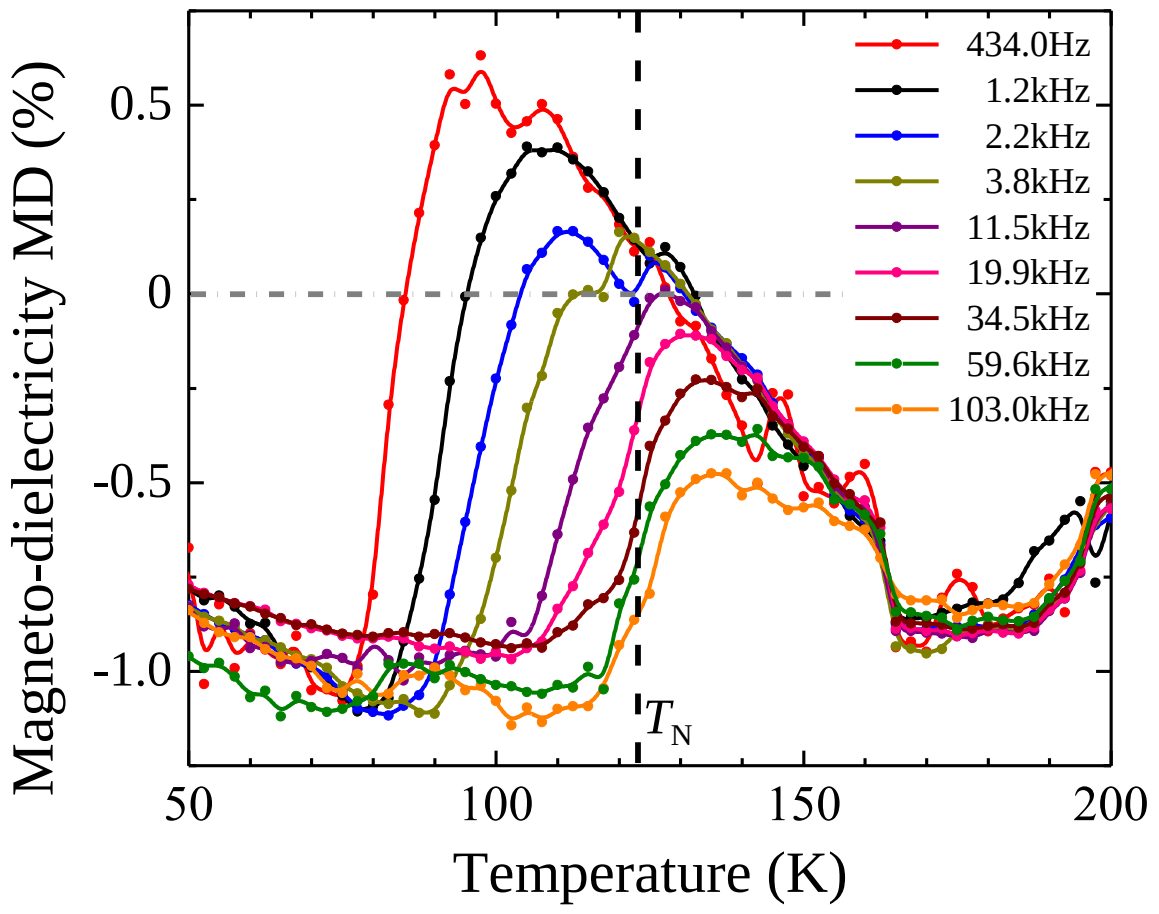

**Figure 3.** Temperature dependence of Magneto-dielectricity MD(%).

To further correlate electric and magnetic properties of the system, magneto-dielectricity (MD) that is change in dielectric properties under 6 Tesla field has been studied.

$$\mathrm{MD}(\%) = \{\varepsilon'(H) \div \varepsilon'(0) - 1\} \times 100 \qquad (2)$$

**Figure 3** demonstrates temperature dependence of MD(%), depicting all negative MD at temperatures above 130K, while at lower temperatures, both negative and positive MD can be seen (for low frequencies). This is already evident in the zero- and 6T-field $\varepsilon'(T)$-plots (figure 2(c)) at selected frequencies. In the system low dimensional magnetic ordering are observed even above 123K, which accounts for MD above $T_N$. Over 80-130K temperature-window, magnetic field suppresses dipole-interactions in larger clusters (c.f., thermal activation), thereby reducing (increasing) the local polarization (polarizability $\varepsilon'$), whereas outside this window, *H*-field consolidates the growth of smaller-sized dipolar clusters (c.f., thermal de-activation), thereby enhancing (reducing) the local polarization ($\varepsilon'$). The dual nature of MD shows up over the temperature region where consolidating dipolar clusters evolve. The exclusive orientational-liquid at high-*T*'s however features only negative MD (i.e., field-enhanced dynamical correlations), as indeed expected.

$R_2$-and $R_1$-relaxations can be characterized as correlated dipoles which consolidate in wake of the emergent/coupled WFM-nanophase and bulk/AFM-matrix. The application of magnetic field tends to merge the AFM transition with occurrence of WFM favoring the emergence of isolated WFM nano-phase (coexistent with the AFM-matrix) versus phase-separated or uniformly spin-canted configurations. With application of 7T magnetic field the AFM-WFM transition merges and shifts down to 109K.[10] Since the WFM is accompanied with octahedron distortion in the system, magneto-dielectricity in the system carries this feature showing both correlation and de-correlation of dipolar entities under the *H*-field, accounting for the positive MD at low-frequencies along with negative MD, over certain temperature window.

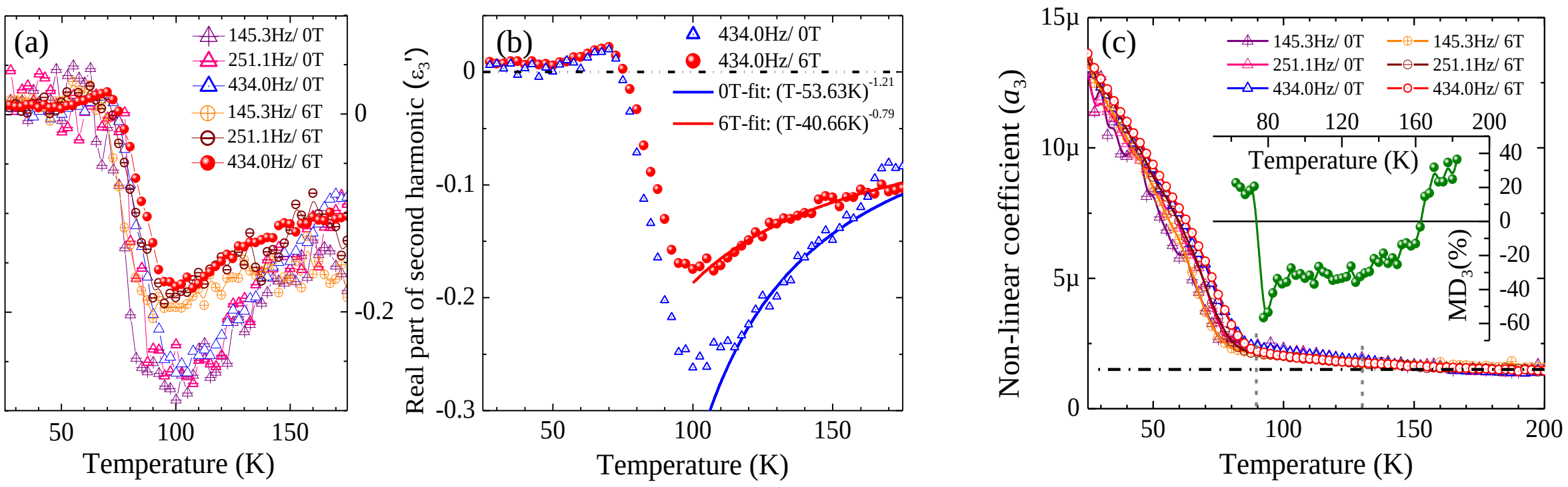

**Figure 4.** (a): Temperature dependence of second harmonic signal ($\varepsilon_3'$) under zero- and 6T-magnetic field. (b): Temperature dependence of $\varepsilon_3'$-signal at 434 Hz with divergence-fit under zero- and 6T-magnetic field. (c): Temperature dependence of $a_3$ signal under zero- and 6T-magnetic field. Inset: Harmonic magneto-dielectricity $MD_3$(%) from the second-harmonic signal.

The magneto-electric effect coupled through structural distortions collaterally manifests in the second-harmonic ($\varepsilon_3'$) measurements. Negative-definite $\varepsilon_3'(T)$-signal has been observed, undergoing through an extremum at ~100K, and vanishing near ~50K, as shown in **figure 4(a)**. This is in agreement with theoretical modeling for relaxor, exhibiting frequency-dispersive relaxation.[19,20] The low strength of $\varepsilon_3'$-signal indicates its major contribution from the term ~ -$\chi_1^4$, denoting cluster-induced relaxor, and not from the bulk polarization term ~ $P^2\chi_1^5$.[20,21]

Increase of the harmonic signal is expected to diverge critically, with temperature dependence[22,23]

$$|\varepsilon_3'| \propto \left(T - T_g\right)^{-\gamma} \tag{3}$$

From the fits of $\varepsilon_3'(T, H)$-signals over 115-160K with equation (3) (solid curves in **figure 4(b)**), we obtain $T_g$(0T) =53.63K as the freezing temperature with divergence exponent $\gamma$(0T) =1.21 at zero-field— comparing well with $\gamma$=1.25 found for BTZ35 by Kleemann et. al.— and $T_g$(6T) =40.66K with $\gamma$(6T) =0.79 at 6T-field.[24]

$$\mathrm{MD}_3(\%) \ = \{\varepsilon_3'(H) \div \varepsilon_3'(0) - 1\} \times 100 \tag{4}$$

It is to be noted that the finite & negative harmonic $MD_3$(%) occurs over roughly the same temperature-window with the positive fundamental MD. Here we can say re-phasing of spatially separated magnetic phases is responsible for the observed harmonic $\varepsilon_3'$-signal.

To further explore the nature of relaxations, non-linear dielectric coefficient ($a_3$) has been evaluated in respective temperature window.[25]

$$a_3 = \ -\varepsilon_3' \div \varepsilon'^4 \tag{5}$$

**Figure 4(c)** shows the obtained dielectric non-linearity where, at high temperatures nearly constant $a_3$ signifies para-electric like region of the system with least correlated dipoles. On decreasing temperature, at ~125K a rise in slope of $a_3(T)$ features relaxor behaviour. On further cooling below 90K a steep increase in $a_3$ manifests growth of more “glass-like” behaviour in the relaxor. Monotonous increase in temperature dependence of $a_3$ from para-electric to relaxor behaviour at ~$T_N$, further featuring glassy-like behaviour in relaxor below $T_{WFM}$ marks non-linear magneto-dielectricity in the system.

Harmonic signal is $H$-dependent and exists ($\varepsilon_3'(T, H) \neq 0$) only down to the freezing temperature ~50K of the segmented dynamics; yielding $MD_3 \neq 0$ over the $T$-window hosting dynamically-activated dipoles. Over 80-160K, $\varepsilon_3'(T)$-signal decreases under the applied $H$-field, giving negative $MD_3$. Expectedly, under-field de-correlation of dipoles relaxing at low-frequencies (yielding MD > 0) also thermally randomizes the 'higher-moments' thereby decreasing the harmonic response.

## Conclusions

We have analyzed the signatures of interacting-dipole structures featuring vitreous relaxation kinetics upon cooling below the AFM transition and the emergence of weak ferromagnetic nano-phase at $T_W$ ~110K, in the AFM matrix ($T_N$ =123K). Alterations of dipoles' correlation-status are allied with the system's magnetic evolution and reflect in the magneto-dielectric signals, upon magnetic field application. In respective temperature regime, spin-lattice coupling has been established in $Ca_3Mn_2O_7$ by analyzing its Raman spectra. Observed second-harmonic magneto-dielectricity is some two orders of magnitude larger, vis-à-vis that obtained in the fundamental response. Structural, magnetic, and electrical coupling in this Ruddlesden-Popper compound is the consequence of its layered structure, which is found to exhibit a topological character. Hence, the ME-coupling can be tailored in its thin-film or nano-fiber formations, where the exclusive features can be enhanced/suppressed by dimensional reduction and intrinsic strain. Furthermore, chemically-doped variants may be investigated to tune the magneto-electric functionality.

## Acknowledgments

Suresh Bhardwaj (UGC-DAE CSR) is thankfully acknowledged for his technical help and facilitation of the dielectric-measurements. We extend our sincere thanks to Mr. Ajay Rathore (UGC-DAE CSR) for help with Raman measurements.